\documentclass[%
 aip
rsi,
 amsmath,amssymb,
reprint,%
]{revtex4-1}

\usepackage{graphicx}
\usepackage{dcolumn}
\usepackage{bm}

\usepackage[utf8]{inputenc}
\usepackage[T1]{fontenc}
\usepackage{mathptmx}

\begin{document}


\title{N\NoCaseChange{b}$_{3}$S\NoCaseChange{n} MULTICELL CAVITY COATING SYSTEM AT JLAB\thanks{Co-Authored by Jefferson Science Associates, LLC under U.S. DOE Contract No. DE-AC05-06OR23177. This material is based upon work supported by the U.S. Department of Energy, Office of Science, Office of Nuclear Physics.}}

\author{G. Eremeev}
 \altaffiliation[now at ]{FNAL}
\email{grigory@fnal.gov}
\author{W. Clemens}%
\author{K. Macha}%
\author{C.E. Reece}%
\author{A.M. Valente-Feliciano}%
\author{S. Williams}%

 \affiliation{ Jefferson Lab, Newport News, VA, USA}%

\author{U. Pudasaini}
\author{M. Kelley}
\affiliation{%
College of William and Mary, Williamsburg, VA, USA
}%

\date{\today}

\begin{abstract}
SRF niobium cavities are the building blocks of modern accelerators for scientific applications. Lower surface resistance, higher fields, and high operating temperatures advance the reach of the future accelerators for scientific discovery as well as potentially enabling cost-effective industrial solutions. We describe the design and performance of an Nb$_{3}$Sn coating system that converts the inner surface of niobium cavities to Nb$_{3}$Sn film. Niobium surface, heated by radiation from the niobium retort, is exposed to Sn and SnCl$_{2}$ vapor during the heat cycle, which results in about 2 $\mu$m Nb$_{3}$Sn film on the niobium surface. Film composition and structure as well as RF properties with 1-cell R\&D cavities and 5-cell practical accelerator cavities are presented.
\end{abstract}

\maketitle

\section{Introduction}
Superconducting RF (SRF) cavities are building blocks of modern particle accelerators. State-of-the-art SRF cavities accelerate particle beams in CEBAF, SNS, European XFEL and other accelerators\cite{Freyberg,SNS,Abela:77248, Reece}. The energy reach and operational efficiency is defined by the properties of the top-most surface layer of SRF cavity surfaces, where surface RF current within about 100 nm shield the superconductor Meissner state. The quality of the niobium surface have been steadily improved over the years by growth in the understanding of the relationship between surface treatments and RF properties of the underlying superconductor\cite{Padamseebook1,Padamseebook2}. While the work to improve niobium surface, which is the superconductor of choice so far, continues, interest is growing towards superconductors with a higher transition temperature. In particular, Nb$_{3}$Sn has been revisited and could be a solution for near-term compact scientific and industrial accelerators due to its potential to sustain a factor of two higher magnetic fields and two orders of magnitude higher quality factor than that of niobium\cite{PosenHall}.
\newline\indent Various technique have been developed to grow Nb$_{3}$Sn superconductor. Bronze routes, sputtering techniques, tin dip et cetera have been studied (see \cite{PosenHall, ValenteRev, Illina, Sayeed}  and references therein), but require further development due to the exceptional quality of the RF surface necessary to sustain high RF fields. The best Nb$_{3}$Sn-coated SRF cavities so far have been produced with the so-called vapor diffusion process\cite{PosenPOP}. The process is attributed to Saur and Wurm\cite{SaurWurm} and comprises exposure of niobium surface to tin vapors at temperature above about 900 $^{\circ}$C. This process has been and is being typically used to grow several $\mu$m-thick Nb$_{3}$Sn layers on the inside of niobium SRF cavities\cite{Siemens, Wuppertal, Cornell, JLab, FNAL}. This contribution describes the developed Nb$_{3}$Sn-coating system and coating process for coating R$\&$D as well as accelerator SRF cavities.

\section{System design}
\subsection{System concept}
The deposition system comprises two main parts: a custom high-vacuum furnace, built by a commercial vendor according to Jefferson Lab specifications, provides a clean heating environment and a high-vacuum coating chamber built at Jefferson lab that contains coated samples and process vapors. System was designed to separate process vapor in the coating chamber from furnace heating environment in order to avoid cross contamination and to increase the flexibility and longitivity of the system. 
\subsection{High vacuum furnace}
The top-loaded vertical furnace was procured from T-M Vacuum Products Inc. The furnace was specified to reach 1315 $^{\circ}$C with the vacuum in 10$^{-7}$ Torr range empty and was built with three hot zones. Each hot zone comprises two 2"-wide low-resistance molybdenum flat elements powered by a 20 kW power supply. Each hot zone is independently controlled by a calibrated molybdenum-sheathed type R thermocouple. Three hot zones create an effective cylindrical heating space of about 50 cm long and 40 cm diameter, where the temperature is uniform to within 3 degree and controlled to within 1 degree by an Allen-Bradly 5/05 programmable logic controller (PLC). Three molybdenum and two stainless steel sheets outside the heating elements serve as radiation shields to isolate the hot zone from the electropolished vacuum vessel, which is water-cooled by 6 gpm room temperature water. The interface door on the top of the furnace was modified to increase process volume. A custom water-cooled retort door was built by Lesker company. A new extension to the existing heat shield with six molybdenum sheets was built in-house to extend the hot zone length to about 80 cm.
 \newline\indent Prior to the process initiation, the chamber is evacuated by a Leybold SC 30 scroll pump. When the pressure drops to below 0.1 mbar, the roughing pump is isolated and shut off. The gate valve to a Sumitomo Marathon 250 cryopump is opened to establish high vacuum in the furnace. Vacuum levels are monitored with thermocouple tubes and a single cold cathode sensor. The standard base vacuum prior to start of the heating process is below 10$^{-8}$ Torr. A typical furnace vacuum level during process runs rises into 10$^{-7}$ Torr range, sometimes reaching into lower 10$^{-6}$ Torr for about 10-20 minutes. Residual gas analysis indicates hydrogen as the dominating specie at high temperatures.
\subsection{Niobium reaction chamber}
The coating chamber was built out of 4 mm niobium sheets into a 17 inch OD by 40 inch long cylinder via rolling half cylinders and electron beam welding them. Furnace heating elements on the outside irradiate the cylinder, heating it to the desired temperature up to 1300 $^{\circ}$C. The cylinder irradiates the inside, where the coated samples are located, heating the samples to the desired temperature. The niobium cylinder is closed from one end with a 4 mm niobium blank, which was deep-drawn into dome shape and electron beam welded to the cylinder shell. The other end of the cylinder shell was welded to a 21" OD titanium (Ti6Al4V) flange with a half-dovetail 1/4 inch o-ring groves for Viton$^{TM}$ o-ring seals. 
 \begin{figure}
\includegraphics*[width=80mm]{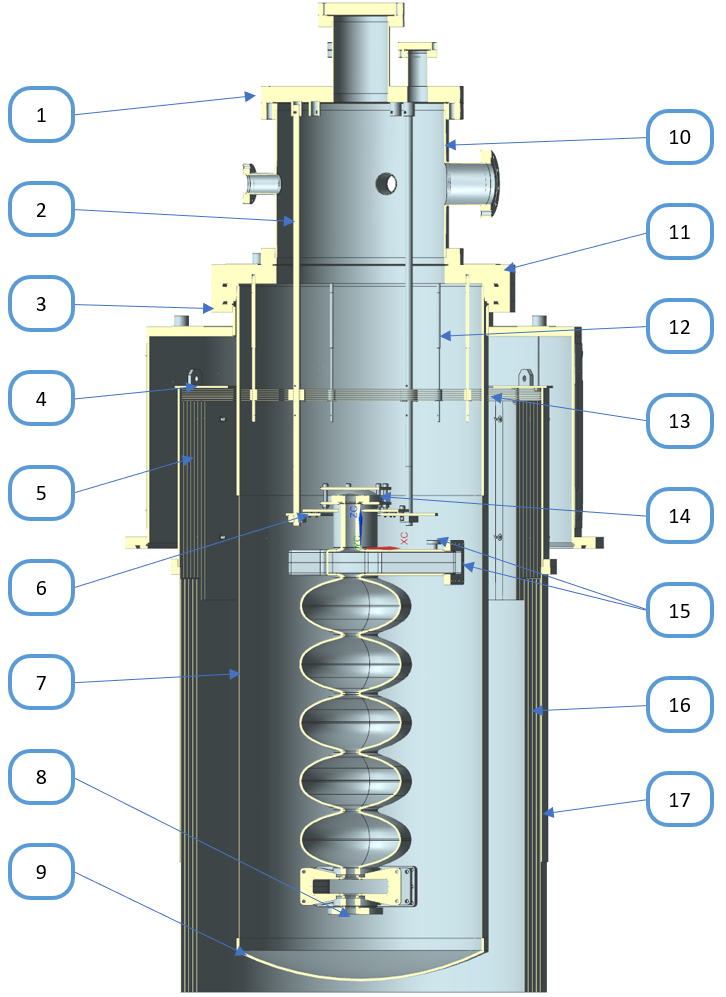}
\caption{\label{Systemdiscr} A schematic of a coating chamber with a CEBAF 5-cell cavity inside the hot zone. 1. Multiport top plate custom-built by Lesker company; 2. 0.5" OD niobium support rods; 3. water-cooled SS door custom-built by Lesker company; 4. SS support structure; 5. Molybdenum heat shields; 6. 4mm-thick niobium support plate; 7. 4mm-thick thick niobium cylinder; 8. Niobium crucible(not shown); 9. 4mm-thick deep-drawn niobium dome, which is electron beam welded to niobium cylinder; 10. Multiport SS spool piece custom-built by Lesker company;  11. Water-cooled zero length reducer custom-built by Lesker company; 12. 1/4" niobium support rods; 13. Niobium and molybdenum heat shields; 14. Niobium cavity support structure; 15. Niobium covers (not shown); 16. Heat shields(a part of the furnace custom-built by T\&M vacuum); 17. SS support structure (a part of the furnace custom-built by T\&M vacuum).}
\end{figure}
On one side of the flange the seal provides vacuum insulation via o-ring to the new furnace door, on the other side the o-ring seals against a zero-length water-cooled reducer, which reduces the opening from 21 inch OD to 14 inch copper gasket seal. This zero-length reducer allows to use instrumentation and pumping on a smaller multiport spool piece and top plate.
\newline\indent The reaction chamber is initially evacuated with either Edwards nXDS 15i pump or Edwards TIC pumping cart to below 10 Torr. Once the pressure reaches 10 Torr, Pfeiffer HiPace 300C turbopump is turned on to bring the pressure to below 10$^{-5}$ Torr before the process is started. Vacuum levels are monitored with Edwards active Pirani vacuum gauges (APG100) for soft vacuum and Edwards Active Inverted Magnetron (AIM) Gauge for high vacuum. Residual gas analyzer (RGA) SRS300 from Stanford Research system is sometimes used to record residual gas composition during the process.
\newline\indent Three 0.5 inch OD niobium rods extend into the hot zone space from the top plate of the multiport spool piece. Six heat shields made out of niobium are attached to the rods to reduce radiative heat loss from the hot zone. A sample chamber or SRF cavities are suspended from a 4 mm niobium plate, which is attached to the rods using molybdenum hardware. Niobium- or tantalum-sheathed type C thermocouples are attached to the coated cavities at several locations in order to monitor the temperature of the reaction chamber during the coating process.
\section{Coating description}
Sn (99.999\% purity from Sigma Aldrich) and SnCl$_2$ (99.99\% purity from Sigma Aldrich) packaged inside niobium foils are placed inside the niobium crucible. About 3 mg/cm$^{2}$ of Sn and similar amount of SnCl2 is used for each Nb$_{3}$Sn coating. Niobium covers, 3 - 4 mm thick, are used to cover open sample chamber or cavity ports and restrict gas flow from the inside of the cavity, where Sn and SnCl2 are placed, during the process. Witness samples are hung inside the cavity by attaching them to the top cover using niobium wires.  The crucible as well as all other niobium covers are assembled to a sample or RF cavity with molybdenum hardware in the cleanroom and double bagged before being transferred to the thin film lab. Typically, molybdenum hardware that is used to fasten niobium covers to open ports is lightly tightened to about 1 ft-lb. In the thin film lab the cavity is attached to the deposition chamber and installed in the Nb$_{3}$Sn deposition chamber. After the furnace and the reaction chamber are evacuated to UHV and HV respectively, heating cycle is started. 
\begin{figure}
\includegraphics*[width=80mm]{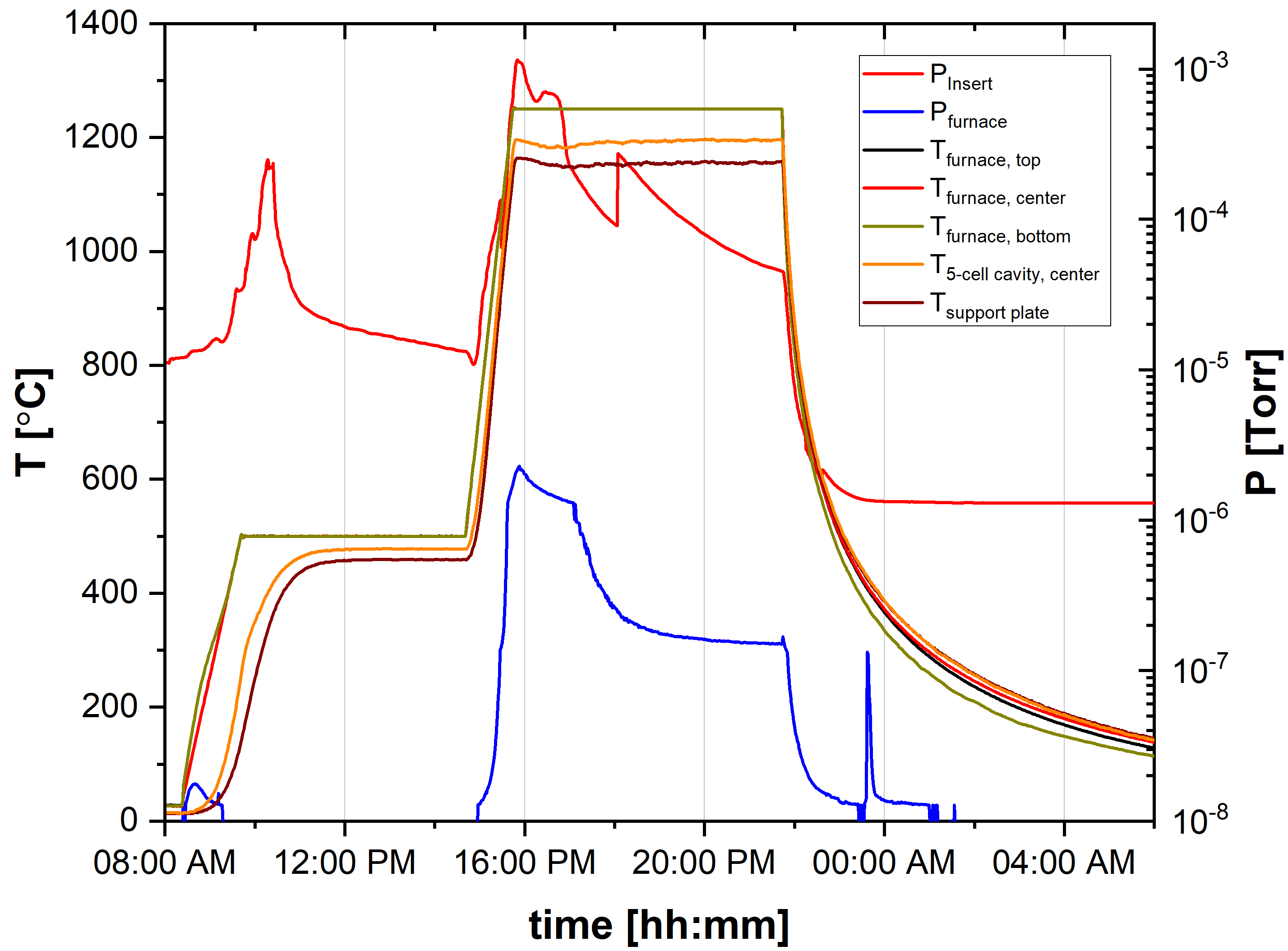}
\caption{\label{Tprocess} Process parameters during the standard coating process. The typical first step is the temperature ramp at 6 $^{\circ}$C/min. The coating chamber is then parked at about 500 $^{\circ}$C for 1-5 hours. The temperature then is ramped at 6 $^{\circ}$C/min to 1200-1250 $^{\circ}$C. Nb$_{3}$Sn layer is then grown at 1200-1250 $^{\circ}$C for 3-24 hours. Note that 5-cell cavity temperature is typically about 50 $^{\circ}$C lower than the furnace control temperature measured outside the niobium cylinder((7) in Fig.$\ref{Systemdiscr}$).}
\end{figure}
The standard temperature profile has two temperature ramps: the first ramp to 500 $^{\circ}$C at 6 $^{\circ}$C/min, and the second ramp to 1200 $^{\circ}$C at 12 $^{\circ}$C/min, Fig. $\ref{Tprocess}$. The standard temperature profile has two plateau regions, at 500 $^{\circ}$C for 1 hour and 1200-1250 $^{\circ}$C for 3-6 hours. During the first temperature plateau at 500 $^{\circ}$C, which is lower than the boiling point of SnCl$_2$ (623 $^{\circ}$C), but is above its melting point of SnCl$_2$ (247 $^{\circ}$C), SnCl$_2$ reacts with the niobium surface, and Sn nucleation sites are formed on the surface. The nucleation produces Sn droplets on the surface from few tens of nanometer up to about half a micron size on the surface\cite{Pudnucl}. By the time the second plateau is reached at 1200 $^{\circ}$C Sn vapor deposits on the surface around the nucleation sites and Nb$_3$Sn grains begin to crystallize on the surface. On the second plateau at 1200 $^{\circ}$C, Nb$_3$Sn grains continue to grow on the surface in the presence of Sn vapor. One hour at 1200 $^{\circ}$C results in the Nb$_3$Sn grains on the order of a few hundred nanometers. Finally, after 3 hours at about 1200 $^{\circ}$C Nb$_{3}$Sn grains grow to 1-2 $\mu$m in size\cite{Pudgrowth}. The temperature profile shows the temperatures measured with the three furnace thermocouples located inside the furnace, but outside the coating chamber. The thermocouples are separated vertically by 10 cm, and typically are within $\pm$ 1 $^{\circ}$C of the target temperature during soaks. Inside the chamber niobium- or tantalum-sheathed type C thermocouples are used to monitor the temperature of the coated samples. In case of large structures such as 5-cell CEBAF cavity temperature measured on the cavity is about 50 $^{\circ}$C lower than the temperature indicated by the control thermocouples outside, cf. orange and gold lines in the Fig. $\ref{Tprocess}$. The vacuum curves shows the vacuum in the furnace, inside and outside the coating chamber, Fig. $\ref{Tprocess}$. The vacuum inside the coating chamber is typically about 10$^{-5}$ Torr, and increases during the coating run up to about 10$^{-3}$ Torr due to Sn and SnCl$_{2}$ vapors.
\newline\indent Single-cell R\&D cavities are coated using both one-cavity and two-cavity setups. The best results are achieved using two-cavity setup. In this setup two cavities are high pressure water rinsed separately and then assembled together at one of the flanges using niobium brackets and molybdenum hardware inside the cleanroom. Two cavities then form the reaction chamber. They are covered with a niobium cover on the top and niobium crucible on the bottom, which hosts 6 g of Sn and 3 g of SnCl$_{2}$ and is covered with molybdenum or niobium diffuser. The best results are achieved when another small tin source is added at the top and a temperature gradient of about 80 $^{\circ}$C, as measured by the thermocouples attached to the cavities, from the top to the bottom is used.
\newline\indent In the case of 5-cell cavities, the bottom crucible is loaded with 10 g of Sn and 3 g of SnCl$_{2}$. The best results are achieved when additional Sn sources are used. In such cases one or two crucibles are added, which are positioned at the top and at the center of the cavity. They are typically loaded with 1-2 g of Sn and the amount of Sn in the bottom crucible is lowered accordingly. Nucleation step in 5-cell cavity coatings is extended to 5 hours and the coating step is typically extended beyond 3 hours. It was 1250 $^{\circ}$ for 6 hours for the cavity coatings reported below.

\section{SRF film performance}
\subsection{Analysis of small samples}
\begin{figure}
\includegraphics*[width=80mm]{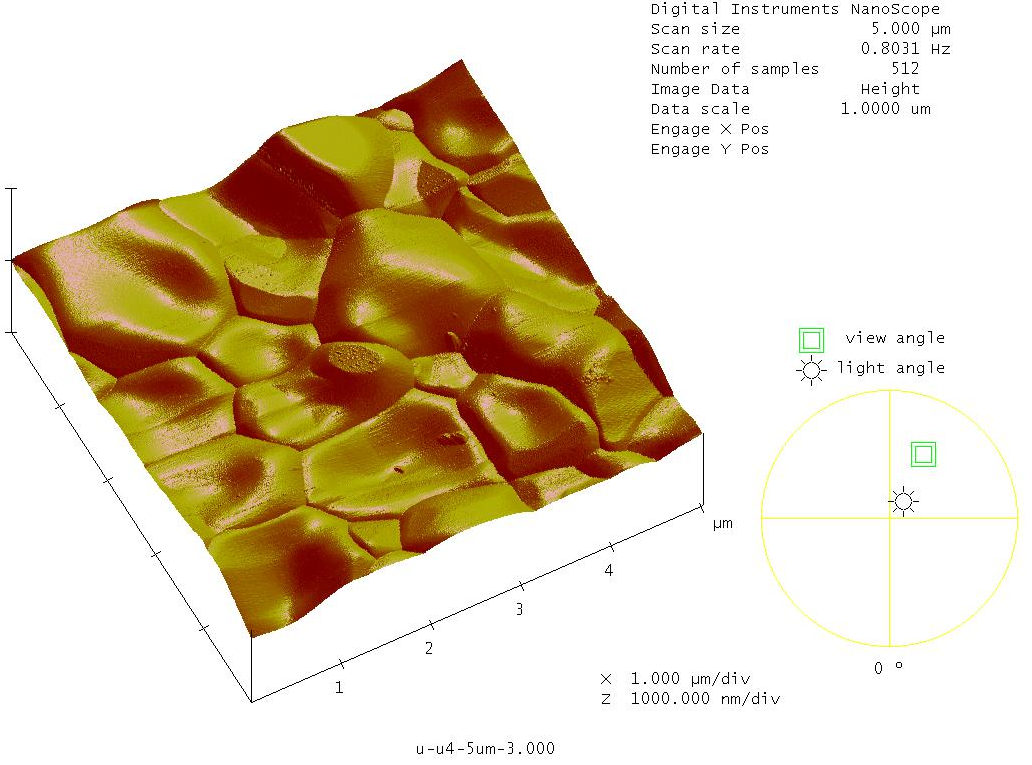}
\includegraphics*[width=80mm]{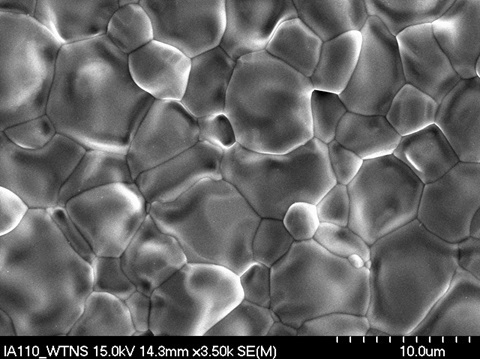}
\caption{\label{AFMfilms} Typical AFM [top] and SEM [bottom] images of coated films are shown. Typical grain size is 2-3 $\mu$m as seen in SEM image. Typical roughness is about 50 nm calculated for 5 x 5 $\mu$m$^{2}$ AFM scans.}
\end{figure}

\begin{figure}
\includegraphics*[width=80mm]{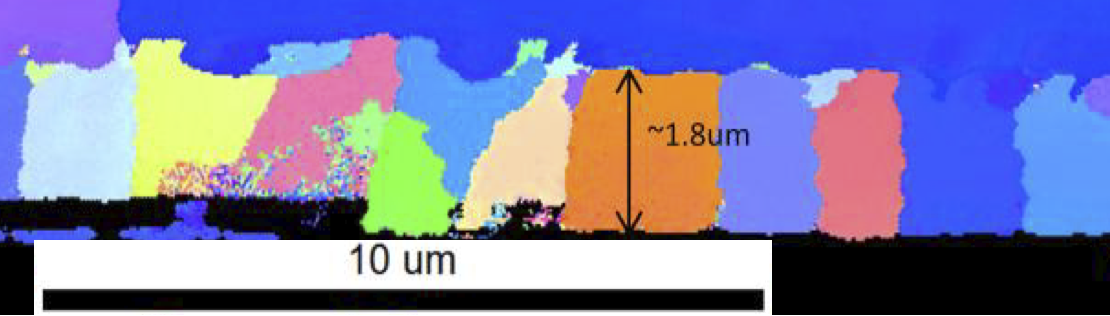}
\caption{\label{EBSDfilms} Typical EBSD image of the crosssection of a coated film is shown. Nb$_{3}$Sn grains are at the bottom and the thickness of the film is indicated with an arrow. Different colors correspond to different orientations of Nb$_{3}$Sn grains.}
\end{figure}

10 cm x 10 cm coupons are cut from 3mm thick high RRR (300) niobium sheet material by EDM cutting. Each sample receives $\approx$ 100 $\mu$m BCP removal using a solution of 49\% HF, 70\% HNO3 and 85\% H3PO4 in the ratio of 1:1:1 by volume. These samples are either coated in a sample coating chamber or as witness samples with RF cavities using the Nb3Sn coating procedure described above. 
\newline\indent The surface topography is measured with with a Digital Instruments Nanoscope IV AFM in tapping mode using silicon tips with diameter less than 10 nm. Each sample was scanned in three different regions with scan size 5 $\mu$m by 5 $\mu$m. AFM images revealed depressions on Nb3Sn grains formed by curved facets, Fig. $\ref{AFMfilms}$. 
These depressions were also visible in SEM images as a dark spot in many grains, see Fig. $\ref{AFMfilms}$. Typical average RMS roughness was found close to 70 nm in each coated sample for 5 $\mu$m by 5 $\mu$m scans\cite{PudTopo}.
\newline\indent EBSD map of the coated sample crosssection is shown in Fig. $\ref{EBSDfilms}$. The map shows that the grains apparent in the SEM image are single crystals. There is no clear preferential grain orientations or correlation to the adjacent grains and underlying niobium. No significant crystal orientation variation inside the grains, nor evidence of lateral or depth compositional variation in the coated Nb3Sn is observed\cite{Tuggle}. The EBSD data is consistent with EDS measurement, which show that compositional variations in our Nb$_3$Sn is within instrument sensitivity and reproducibility (24.3 $\pm$ 2) at. \% .
We have not found significant variations in the Nb$_3$Sn composition outside this range. SIMS data was collected on some samples with a CAMECA IMS-7f magnetic sector instrument. The data shows that Nb$_{3}$Sn extends to the depth of about 1 $\mu$m with little variation in Nb and Sn concentrations. SIMS results were further corroborated by XPS data, which was collected on ULVAC-PHI ”Quantera SXM”. High resolution XPS scan of Nb and Sn peaks shows that Nb3Sn is covered with Nb$_2$O$_5$ and SnO$_2$.
\newline\indent The transition temperature of one of the coated samples was measured via 4-probe resistive measurement technique. Transition temperature was found to be 18 K, which is consistent with the transition temperature from cavity measurements.
\subsection{Single-cell cavity results}
\begin{figure}
\includegraphics*[width=80mm]{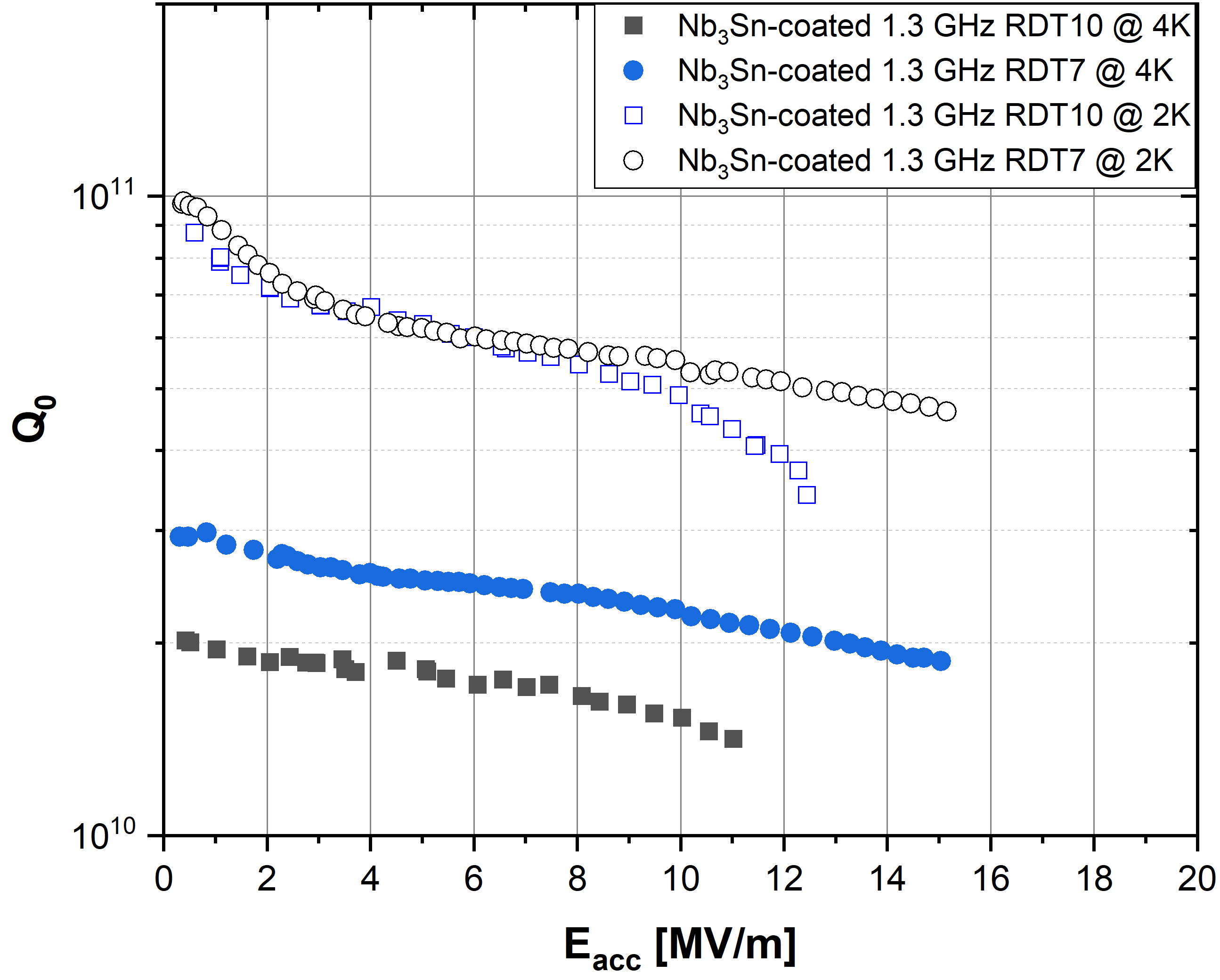}
\caption{\label{1cellresults} Single cell cavity results, cf. typical Q$_{0}$ before Nb$_{3}$Sn coating at 4K is 3$\cdot$10$^{8}$, and Q$_{0} $ at 2K was 2$\cdot$10$^{10}$. }
\end{figure}

Single-cell cavities RDT7 and RDT10 were made from high purity (RRR $\approx$ 300) fine-grain niobium sheet by stamping and electron beam welding. Each cavity has received the standard surface preparation after manufacturing: over 100 $\mu$m surface material removal and several heat treatments at 800 $^{\circ}$C for two or three hours\cite{Geng1}. In the baseline test before Nb$_{3}$Sn coating the cavities reached the accelerating gradient (E$_{acc}$) of about 30 MV/m with the low-field Q$_0$ of about 1.6$\cdot$10$^{10}$ at 2K. 
\newline\indent Cavities were coated using the two-cavity setup.  A temperature gradient of about 85 $^{\circ}$C between the top (cold) and bottom (hot) of paired cavity setup was used. Almost Q-slope free results were achieved as shown in Figure $\ref{1cellresults}$. The measured value of RDT7 low-field Q0 was 3$\cdot$10$^{10}$ at 4 K and 10$^{11}$ at 2 K without a significant Q-slope. The cavity maintained a Q$_0$ of about 2$\cdot$10$^{10}$ at 4 K and above 4$\cdot$10$^{10}$ at 2 K before quench at above 15 MV/m. Post-coating inspection showed uniform coating inside the cavity. Tin consumption was very similar between the two coatings. Examination of witness samples showed uniform coating without any tin residue or patches\cite{PudUniform}.  EDS examination showed usual Nb$_{3}$Sn composition. RF test results from RDT10 were similar to RDT7 except for the quench field, which was lower in RDT10. 
Coated cavities at 4K exhibited quality factor of about 3$\cdot$10$^{10}$, which is almost two orders of magnitude higher than that of niobium cavities. Field-dependence of surface resistance caused quality factor degradation to 2$\cdot$10$^{10}$ at E$_{acc}\approx$15 MV/m, equivalent to peak surface magnetic field of about 60 mT. Field reach was limited by a thermal breakdown, which has been a common limitation of Nb$_{3}$Sn cavity coated by vapor diffusion. At lower temperature of 2 K low field quality factor reached 10$^{11}$ corresponding to about 2-3 nOhm of surface resistance similar to the lowest observed in SRF Nb$_{3}$Sn-coated cavities.  Quality factor exhibited a non-trivial field-dependence quickly reducing to about 7$\cdot$10$^{10}$, corresponding to about 4 nOhm at E$_{acc}$=2 MV/m (8 mT), and reaching 4.5$\cdot$10$^{10}$ (6nOhm) at about E$_{acc}$=16 MV/m (70 mT). The surface resistance of these cavities is one of the best measured at 60-70 mT to date.
\subsection{Multicell cavity performance}
Multicell cavity performance historically trailed single-cell R$\&$D results\cite{Geng}. Possible reason is the higher complexity due to the large surface area of multicell cavities and addition of auxiliary components, such as HOM couples, which complicates application of uniform surface treatment. The coating system was designed to coat CEBAF 5-cell cavities. Two 5-cell cavities, 5C75-RI-NbSn1 and 5C75-RI-NbSn2, were built by RI Research Instrument, GmbH according to Jefferson Lab's specifications. The cavities were built to C75 shape with HOM and FPC coupler, which allows these cavities to be integrated into a CEBAF cryomodule. The cavities were electropolished for 120 $\mu$m removal, annealed in vacuum at 800 $^{\circ}$C for two hours, and were again electropolished for 25 $\mu$m as the final material removal step. In the baseline test 5C75-RI-NbSn1 reached E$_{acc}\approx$29 MV/m with the low-field Q$_0$ of about 2$\cdot$10$^{10}$. 5C75-RI-NbSn2 was measured up to E$_{acc}\approx$22 MV/m with the low-field Q$_0$ of about 2$\cdot$10$^{10}$. 
\newline\indent Both cavities were coated using the 5-cell cavities setup described above. After Nb$_{3}$Sn coating cavities were ultrasonically cleaned, rinsed with high pressure (80 bar) ultra pure water, assembled, evacuated and tested. Both cavities were tested in liquid helium bath at 4.4 and 2.0 K. 
\begin{figure}
\includegraphics*[width=80mm]{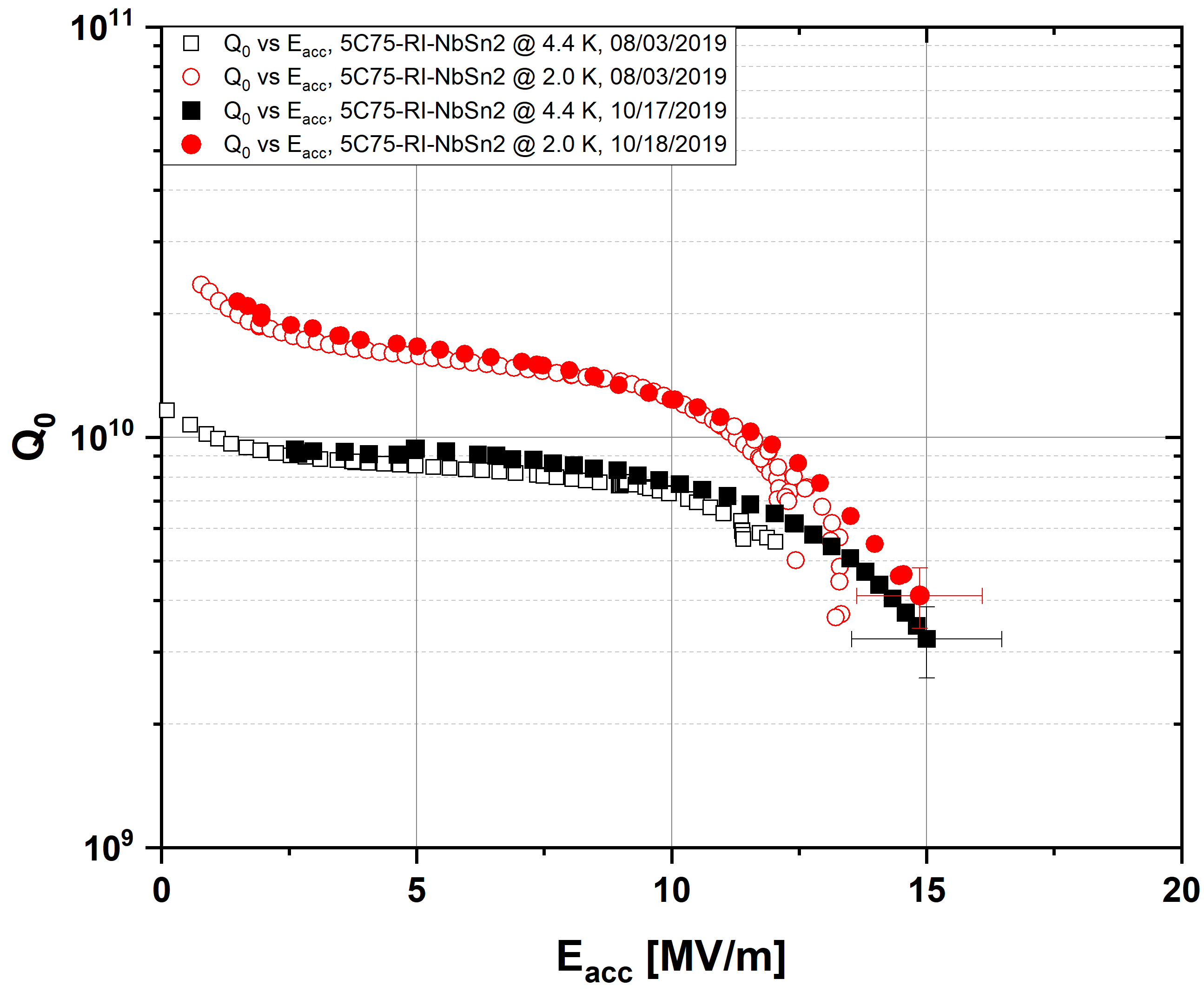}
\caption{\label{5cellresults} Multicell cavity results. Note Q$_{0}$ close to 10$^{10}$ at 4K, which is a about a factor of 30 improvement over a typical uncoated niobium cavity.}
\end{figure}
Non-uniformity was observed in the coating of 5C75-RI-NbSn1, which was linked to low tin pressure in the early stages of the coating. To improve tin availability, three tin sources were used to coat 5C75-RI-NbSn2 in configuration described above. The low-field quality factor was about 1.2$\cdot$10$^{10}$ corresponding to about 20 nOhm of surface resistance at 4.4 K. At 2 K the quality factor improved to about 2$\cdot$10$^{10}$ at low fields.  Field dependence of the surface resistance caused the reduction in the quality factor with field, but, because X-rays were observed in the measurements, it was not clear whether the field dependence was inherent to the film or was caused by field emission loading. The cavity was then shipped to FNAL, where it was partially disassembled, rinsed with high pressure ultra pure water, assembled, evacuated and tested. The test at FNAL was carried out about two month after the test at JLab. The low-field quality factor was measured almost exactly the same to those measured at JLab at both 4.4 K and 2 K, Fig.$\ref{5cellresults}$. Field-dependence of the surface resistance was still present and caused the reduction in the quality factor to about 3$\cdot$10$^9$ (90 nOhm) at about 15 MV/m (60 mT) at 4.4 K. No X-rays were observed in the latest measurement.  
\section{Conclusion}
We have designed and commissioned a coating system, comprising a commercial furnace and custom niobium reaction chamber, for coating Nb$_{3}$Sn films on the inner surface of SRF cavities. The system is capable of coating multicell accelerator cavities inside niobium reaction chamber with vacuum system separate from the furnace vacuum. 2 $\mu$m thick uniform Nb$_{3}$Sn films with the transition temperature of 18 K are grown on small samples and multicell cavities.
\newline\indent Coated single cell cavities exhibit quality factors close to 3$\cdot$10$^{10}$ at 4 K, about two orders of magnitude improvement over typical quality factor, and close to 5$\cdot$10$^{10}$ at 2 K, a factor of 3 improvement of baseline tests of these cavities. Multicell cavities exhibit quality factors in excess of 10$^{10}$ and reach above 10 MV/m, suitable for accelerator applications.
\begin{acknowledgments}
We thank JLab \& FNAL technical staff for help with some of the cavity preparation and Brian Carpenter, Kirk Davis, John Fischer, Tony Reilly, Bob Rimmer for useful suggestions.
\newline\indent Co-Authored by Jefferson Science Associates, LLC under U.S. DOE Contract No. DE-AC05-06OR23177. This mate- rial is based upon work supported by the U.S. Department of Energy, Office of Science, Office of Nuclear Physics.
\end{acknowledgments}

\end{document}